\documentclass[11pt]{article}
\usepackage{latexsym}
\def\cB{{\cal B}}

\def\cF{{\cal F}}

\def\cO{{\cal O}}

\def\cT{{\cal T}}

\newfont{\goth}{eufm10 scaled \magstep1}

\def\a{\alpha}

\def\b{\beta}

\def\c{\gamma}
\def\d{\delta}
\def\e{\epsilon}

\def\k{\kappa}
\def\l{\lambda}

\def\s{\sigma}
\def\t{\tau}

\def\beq{\begin{equation}}\def\eeq{\end{equation}}
\def\beqa{\begin{eqnarray}}\def\eeqa{\end{eqnarray}}
\def\barr{\begin{array}}\def\earr{\end{array}}

\def\del{\partial}

\def \la {\langle}
\def \ra {\rangle}

\def \Z {{\bf Z}}
\def \del {\partial}
\def \dels {\partial\kern-.5em / \kern.5em}
\def \As {{A\kern-.5em / \kern.5em}}
\def \Ds {D\kern-.7em / \kern.5em}

\def \a {\alpha}
\def \b {\beta}
\def \dag {\dagger}

\def \G {\Gamma}
\def \d {\delta}
\def \eps {\epsilon}

\def \k {\kappa}

\def \s {\sigma}

\def \t {\tau}

\def \II {I\hspace{-.1em}I\hspace{.1em}}
\def \IIA {\mbox{\II A\hspace{.2em}}}
\def \IIB {\mbox{\II B\hspace{.2em}}}

\def\gs{\geq}
\def\ur{uncertainty relation }

\let\bm=\bibitem

\def\nn{\nonumber}
\def\bd{\begin{document}}
\def\ed{\end{document}}
\def\ba{\begin{array}}
\def\ea{\end{array}}
\def\bea{\begin{eqnarray}}
\def\eea{\end{eqnarray}}
\def\ft#1#2{{\textstyle{{\scriptstyle #1}\over {\scriptstyle #2}}}}
\def\fft#1#2{{#1 \over #2}}
\newcommand{\be}{\begin{equation}}
\newcommand{\ee}{\end{equation}}
\newcommand{\eq}[1]{(\ref{#1})}
\def\eqs#1#2{(\ref{#1}-\ref{#2})}
\def\det{{\rm det\,}}
\def\tr{{\rm tr}}
\newcommand{\ho}[1]{$\, ^{#1}$}
\newcommand{\hoch}[1]{$\, ^{#1}$}
\def\ra{\rightarrow}
\def\uha{{\hat {\underline{\a}} }}
\def\uhc{{\hat {\underline{\c}} }}
\def\la{\langle}
\def\ra{\rangle}

\def \G{\Gamma}
\def \YM{\mbox{\tiny YM}}

\begin{document}

\hfill{NEIP-99-006}

\hfill{hep-th/9904133}

\vspace{20pt}

\begin{center}

{\Large\bf Worldvolume Uncertainty Relations for D-Branes}
\vspace{30pt}

{\large Chong-Sun Chu\hoch{1}, Pei-Ming Ho\hoch{2} 
and Yeong-Chuan Kao\hoch{2}}

\vspace{15pt}

\begin{itemize}
\item[$^1$] {\small \em Institute of Physics,
University of Neuch\^atel, CH-2000 Neuch\^atel, Switzerland} 
\item[$^2$] {\small \em Department of Physics,  National Taiwan
University, Taipei 10764, R.O.C.}
\end{itemize}

\vspace{60pt}

{\bf Abstract}
\end{center}

By quantizing an open string ending on a D-brane in a 
nontrivial supergravity background, we argue that there is
a new kind of uncertainty relation on a D-brane worldvolume.
Furthermore, we fix the form of the uncertainty relations and
their dependence on the string coupling constant by
requiring them to be consistent with various
string theory and M theory dualities. 
In this way we find a web of uncertainties of spacetime
for all kinds of brane probes,
including fundamental strings, D-branes of all dimensions
as well as M theory membranes and fivebranes.

\pagebreak

\section{Introduction}

String theory is a promising candidate of quantum gravity and the
theory of spacetime. It is  therefore an interesting question to study
the properties of spacetime in string theory. 
A number of uncertainty relations have been proposed in relation to  
string theory.  See, for example, \cite{douglas,liyon1} 
for a review of the subject and for further references.
It was first proposed \cite{early} 
that a modified version of the canonical Heisenberg uncertainty relation
\be
\d X\d P\geq 1+l_s^2\d P^2,
\ee
governs the high energy behavior of string theory. 
This relation implies 
the existence of  a minimal scale $\d X\sim l_s$ \cite{early}.
Another uncertainty relation is the one proposed by 
Yoneya \cite{Yon}
\be \label{TX}
\d T\d X\geq l_s^2.
\ee
He further suggested that the spacetime uncertainty relation
can be one of the fundamental principles underlying
the nonperturbative string theory,
and can also be one of the guiding principles for constructing
a covariant formulation of M theory \cite{Yon}.

In this paper, we propose a new kind of uncertainty relation
\be \label{ours}
\d X^0\d X^1\cdots\d X^p\geq g_s l_s^{p+1}
\ee
for the worldvolume of a D$p$-brane. Here $X^i$ are the D-brane
worldvolume coordinates. We also propose a similar kind of
uncertainty relations for the M-branes. 

It is well known that in string theory different probes 
could
see slightly different spacetime geometries.
So it is natural to expect that the uncertainty principle
can be different for different probes, as it is manifest in
our proposed relations (\ref{ours}).
Note that (\ref{ours}) is involved only with the longitudinal
directions on a D-branes, that is, the D-brane worldvolume.
Another uncertainty relation involving both longitudinal
and transverse directions for a D-brane was proposed
by Yoneya
and collaborators
\cite{liyon1,Yon,yoneya}.
He proposed that if one interprets $T$ and $X$
as the longitudinal and transverse coordinates of a D-brane,
(\ref{TX}) can also be understood as an uncertainty relation
for a D-brane.

In the brane world scenario \cite{bw},
it is argued that we could be living on a D-brane.
It is thus of great interest to determine
the form of uncertainty relations on a D-brane worldvolume,
which will be interpreted as the uncertainty relation
for the four dimensional spacetime on which we are trapped.
Since it is also known that a D-brane worldvolume becomes
noncommutative \cite{CDS,DH} in the presence of a nontrivial background
$B_{NS-NS}$ field in the direction of the brane,
it is natural to ask what kind of uncertainty relation follows
this worldvolume noncommutativity \cite{SJ,CH}.
In this paper we first show how to obtain 
the uncertainty relation for a D1-brane
by integrating out quantum fluctuations of the background field. 
Then we use various dualities in string theory and M theory
to extend the uncertainty relation to other D-branes and
strings or membranes.
Putting all the old and new dualities together, we find a web
of spacetime uncertainty relations associated with
all the branes in string theory or M theory.

The organization of this paper is as follow. In sec.\ref{open}, 
we first extend the previous results \cite{CDS,DH,CH} 
about the noncommutativity of D-brane by generalizing the background to
the first nontrivial (yet manageable) order, with a curved background
metric $g_{\mu \nu}$ and a nontrivial NS-NS 2-form field. 
Then we argue in sec.\ref{noncomm} that these commutation relations
give rise to uncertainty relations on the D-brane worldvolume. 
While the precise form of the uncertainty relations 
cannot be easily fixed this way,  
one can use the dualities of string theory and M theory to constrain
the form of the uncertainty relations.  
Guided by this idea, we propose in sec.\ref{ur} 
worldvolume uncertainty relations for D-branes,
M2-branes, M5-branes, as well as fundamental strings.
We check that they are consistent with
various dualities of string theory and M theory. 
Finally we make a few remarks in sec.\ref{remarks}.

\section{Noncommutativity from open string quantization}\label{open}

It was first proposed by Connes, Douglas and Schwarz \cite{CDS},
and by Douglas and Hull \cite{DH}, that the Matrix model
compactified on a torus with the NS-NS $B$ field background
should be described by a field theory living on a noncommutative space.
This conjecture has been verified in various ways 
(see, e.g. \cite{NCT}).
In particular, it was pointed out in \cite{AAS} that this can be seen
by quantizing an open string ending on a D-brane.
Based on the intimate relation between compactified Matrix models
and D-brane worldvolume theories, Hauffman and Verlinde \cite{HV}
also suggested that the low energy effective theory
for a D-brane compactified on a torus with $B$ field background
should also live on a noncommutative space.
Following \cite{AAS}, the quantization of the open string ending on a D-brane
with a constant $\cF=B-F$ field background
in flat spacetime was carried out in \cite{CH}.
An agreement with previous works was obtained but the framework
and results of \cite{CH} were more general.
In this section we generalize further the previous work
to a curved background and non-constant $\cF$ field.
The derivation is in parallel with \cite{CH}.

For an open string with an end-point sticking
to a D$p$-brane with $U(1)$ field strength $F$ in
a $B$-field background, the bosonic part of its action
can be written as
\be \label{action}
S_B= {1 \over 4\pi\alpha'} \int_{\Sigma} d^2\sigma
\left[ \eta^{\a\b}g_{\mu\nu} \del_\a X^{\mu}\del_\b X^{\nu}+
\e^{\a\b} \cF_{\mu\nu}\del_\a X^{\mu}\del_\b X^{\nu} \right],
\ee
where
\be
\cF=B-F
\ee
is the modified Born-Infeld field strength on the D-brane,
$F=dA$ and $A_i,\ i=0,1,\cdots, p$, is the $U(1)$ gauge potential
living on the D$p$-brane.
We use the convention $\eta^{\a\b}=\mbox{diag}(-1,1)$
and $\eps^{01}=1$ as in \cite{CH}.
Note that $\cF$ is invariant under the gauge transformations
\be
A\rightarrow A+d\Lambda, \quad B\rightarrow B,
\ee
and 
\be
B\rightarrow B+d\Lambda, \quad A\rightarrow A+\Lambda.
\ee

The equations of motion are
\be \label{EOM}
g_{\mu \k} (\ddot{X}^\k + \Gamma^\k{}_{\mu \nu} \dot{X^\mu}
\dot{X^\nu}) -
g_{\mu \k} ({X''}^\k + \Gamma^\k{}_{\mu \nu} {X'}^\mu
{X'}^\nu) + H_{\mu\nu\l}\dot{X}^{\nu}{X'}^{\l}=0,
\ee
where
\be
{\G^{\l}}_{\mu\nu}=\frac{1}{2}g^{\l\k}
(\del_{\mu}g_{\k\nu}+\del_{\nu}g_{\mu\k}-\del_{\k}g_{\mu\nu})
\ee
is the Riemannian connection for the metric and
\be
H_{\mu\nu\l}=\del_{\mu}\cF_{\nu\l}+\del_{\nu}\cF_{\l\mu}+
\del_{\l}\cF_{\mu\nu},
\ee
that is, $H=d\cF$. In fact, since $dF=0$, $H=dB$.
The boundary conditions are
\be \label{BC}
X^{'\nu} g_{\nu i}  +\dot{X}^{\nu} \cF_{\nu i} =0
\quad \mbox{for} \;\; i=0,1,\cdots,p,
\ee
\be
X^a=x_0^a 
\quad \mbox{for} \;\; a=p+1,\cdots,9,
\ee
at $\s=0,\pi$, where $x_0^a$ gives the position of the D$p$-brane.

For convenience we have chosen the spacetime coordinates
in such a way that $x_0^a$ are constant on the D$p$-brane.
We will use the indices $(i,j,\cdots)$ for directions
parallel to the brane ($i=0,1,\cdots,p$) and the indices
$(a,b,\cdots)$ for directions transverse to the brane
($a=p+1,\cdots,9$).
We will choose the coordinates such that $g_{ia}=0$ 
on the D-brane for convenience.
The momentum density is
\be
2 \pi \a' P_{\mu}= \dot{X}^{\nu}g_{\nu\mu} +X^{'\nu}\cF_{\nu\mu} .
\ee

For a constant background ($\cF=$constant) in a flat spacetime
($g_{\mu\nu}=\eta_{\mu\nu}$),
one can solve the equations of motion and boundary conditions
exactly and carry out the canonical quantization \cite{CH}.
The final result is
\be \label{XX1}
[X^i,X^j]= \pm 2\pi i\a'(M^{-1}\cF)^{ij},
\ee
where
\be \label{M}
M_{ij}=\eta_{ij}-\cF_i{}^k\cF_{kj},
\ee
and $X^i$ is the spacetime coordinates of the open string at
the two end-points $\s=0,\pi$.
The indices are lowered or raised by the spacetime metric $\eta$.
While this result does not need a compactification,
in case the spacetime is compactified on a torus,
the right hand side of (\ref{XX1}) for $F=0$ is
proportional to the dual $B$ field on the dual torus.
It agrees with the results for the BFSS matrix model \cite{CDS}.
For the case of a non-constant $\cF$ on a flat torus and its
relation to deformation quantization, see \cite{GC}.

For a generic background, we are unable to find the most general
solution in parallel of \eq{XX1}. 
However it is possible to find a special solution in 
a certain approximation of weak field and slow variation. 
Consider a deviation from the trivial background $\cF=0$, $g =\eta$
with small $\cF$ and $\del g$ considered as first order quantities.
We will
consider $\del \del g$ as second order and so from the Einstein
equation for the background,
$\del \cF \sim \del g$ are also of the first order.  
In summary, we will consider $(x, \dot{x}, g)$ as terms of the 0th order,
$(\cF, \del\cF, \del g)$ as terms of the 1st order,
and $( \del^{n}\cF,  \del^{n} g), n\geq 2$ as terms of higher orders.
We will only keep terms of the 0th and 1st order consistently.
To solve the equations of motion and boundary conditions,
we use the following ansatz for $X$
\be \label{ansatz}
X^{i}=x^{i}(\t)+y^{i}(\t)\s, \quad X^{a}=x_0^a.
\ee
It is easy to check that this ansatz for the $\s$ dependence 
is consistent with our approximation.
Substituting \eq{ansatz} into the boundary conditions \eq{BC}, 
one finds that $y$ is of first order:
\be \label{BCy}
y^{i}=-\dot{x}^j (\cF g^{-1}(x))_j{}^i.
\ee
For this ansatz, the equations of motion (\ref{EOM})
give the geodesic equation for $x^i$
\be
\ddot{x}^{i}+{\G (x)^{i}}_{jk}\dot{x}^{j}\dot{x}^{k}=0, 
\ee
and the free motion for $y$
\be \label{y}
\ddot{y}^i=0.
\ee 
Eqn.(\ref{y}) is satisfied by (\ref{BCy}) up to 2nd order terms.

The lowest energy modes on a string are thus approximately given by
\bea
X^{i}&=&x^{i}+ (g^{-1} \cF g^{-1}(x))^{ik} p_k \s, \nn \\
2\pi \a' P_{i}&=& p_i - p_m p_n g^{mj} g^{nk}
(\del_j\cF_{ki}(x)) \s,
\label{XP}
\eea
and
\be 
X^a=x_0^a, \quad P_a=0,
\ee
where we have introduced
\be
p_k =  \dot{x}^j g_{jk}(x).
\ee


The Poisson bracket for the open string is determined by
the symplectic two-form
\footnote{
We don't need to do the time average prescription here as in \cite{CH};
and if we do it here we will obtain the same result.
}
\be \label{Omusu}
\Omega=  \int_0^{\pi}d\s\d P_{\mu}\d X^{\mu}.
\ee

Substituting \eq{XP} into (\ref{Omusu}),
we find
\be\label{Om0}
\Omega = \frac{1}{2\a'} \d\bar{p}_{i} \d\bar{x}^i. 
\ee
up to second order terms.
Here $\bar{p}_i$ and $\bar{x}^i$ are defined as
\bea
\bar{p}_i &=& p_i  - \frac{\pi}{2} p_m p_n g^{mj} g^{nk}
(\del_j\cF_{ki}(x)), \\
\bar{x}^i &=& x^i + \frac{\pi}{2}  (g^{-1} \cF g^{-1}(x))^{ik} p_k.
\eea
We thus obtain the commutation relations \footnote{
Normally, special attention has to be paid to the operator ordering 
when we derive commutation relations from Poisson brackets,
so that the Jacobi identity is satisfied if possible. 
Fortunately, in the approximation we use here,
the issue of operator ordering does not arise.
}
\bea
&[\bar{p}_i,\bar{p}_j]=0 , \quad [\bar{x}^i,\bar{x}^j]= 0, \label{xxpp}\\
&[\bar{x}^i,\bar{p}_j]=2 i\a' \d^i_j, \label{xbarp}
\eea
or equivalently, in terms of $x^i$ and $p_j$, it is
\bea
&[p_i,p_j]= 0, \label{pp} \\
&[x^i,x^j]= 2 \pi i \a' (g^{-1}\cF g^{-1})^{ij}, \label{xx} \\
&[x^{i},p_{j}]=2 i\a' \delta^i_j +
i \pi \a' p_k g^{km} g^{in} (\del_m \cF_{nj} + \del_n \cF_{mj}
-\del_j \cF_{mn}). \label{xp} 
\eea
All of these hold only up to 2nd order terms.
It is easy to verify  that the commutation relations (\ref{pp})-(\ref{xp})
satisfy the Jacobi identity also up to second order terms. 


It is now easy to check that for $\s =0, \pi$,
\be
[X^{i},X^{j}]=\pm 2 \pi i \a' (g^{-1}\cF g^{-1})^{ij},
\label{xx1}
\ee
where the plus (minus) sign corresponds to the end-point $\s=0$ $(\pi)$.
This agrees with the result for a constant background (\ref{XX1})
in the leading order.
Furthermore, since the right hand side of (\ref{xx1}) is
a tensor field, this equation is covariant under
general coordinate transformations up to second order terms.

In the static gauge $X^{i}$ is the worldvolume coordinate
for the D-brane, so the D-brane worldvolume appears to be
a noncommutative space.
We propose that in the weak field and slow variation approximation 
we considered,
\eq{xx1} gives the commutation relations for
the D-brane spacetime coordinates in a generic background.

Under an S-duality transformation,
a D-string is turned into a fundamental string. 
For a D-string, the noncommutativity is governed by (\ref{xx1})
with the NS-NS background $\cF= B - dA$.
The commutation relation for a fundamental string
in the dual theory is thus
\be \label{F1CR}
[X^{i},X^{j}]=\pm 2 \pi i g_s' \a' (g^{-1}\cF' g^{-1})^{ij},
\ee
where $g_s'= 1/g_s$ is the dual string coupling and 
$\cF'= B'-dA'$ is the R-R counterpart of $\cF$ in the dual theory.
Since we assumed that $g_s$ is small to derive (\ref{xx1}),
(\ref{F1CR}) is valid only for large $g_s'$.
It would be interesting if one can derive this directly
from string theory.

\section{Noncommutative gauge theory and uncertainty relations}
\label{noncomm}

The noncommutativity \eq{xx1} signifies the existence of
uncertainty relations on the brane. To give a precise
formulation, we first need to explain what we mean by an 
uncertainty $\d X$. 
The suitable framework for discussing uncertainty relation
is to employ the language of  string  field theory \cite{ST}.
Let $\Psi$ be the normalized wave function for the D-brane and define
$(\Delta X^i)^2$ by
\be
(\Delta X^i)^2=\int[DX(\xi)]\Psi^{\dag}(X(\xi))
(X^i(\xi)-\overline{X^i} \; )^2\Psi(X(\xi)),
\ee 
where
\be
\overline{X^i}
=\int[DX(\xi)]\Psi^{\dag}(X(\xi))X^i(\xi)\Psi(X(\xi)).
\ee
In these equations, $\xi=(\xi_0,\cdots,\xi_p)$ denote
the worldvolume coordinates of the D-brane.
The part $[DX(\xi)]$ of the functional measure denotes
an integration over all D-brane configurations.
Obviously these definitions mimic their counterparts in
the familiar case of the quantum mechanics for a point particle.

Applying the standard argument for uncertainty principle to \eq{xx1} 
and notice that as a background, $\cF $ is independent of the D-brane
wavefunction $\Psi$, we get
\be \label{tt}
\Delta X^i \Delta X^j \geq 2\pi l_s^2 |\cF^{ij}|, 
\ee
where $\cF^{ij}=(g^{-1}\cF g^{-1})^{ij}$ and $l_s^2=\a'$.
Note that these first quantized quantities depend on $\Psi$ 
and the 
classical 
backgrounds. The right hand side of \eq{tt} vanishes
for a trivial classical background $\cF=0$. 
However, as explained in \cite{CH}, even for such a 
classical
background, 
there could be nontrivial uncertainty relations arising from 
integrating out the quantum fluctuations. 
On expanding the string field $\Psi$ into the component fields $\cB$, 
the string field path integral becomes an infinite product 
of path integrals over these fields  
and the expectation value $\la \cdot \ra$
of an operator in string field theory is defined by 
\be \la \cO \ra= {1 \over Z}
\int[D\cB] e^{-S} \: \cO
\quad \mbox{where}\quad
Z =\int[D\cB] e^{-S}.
\ee 
For example, $\cB$ includes the metric $g$ and the $B$-field.
It is thus natural to define the desired uncertainty for $X^i$ as
\be \label{def}
\d X^i= \sqrt{\la (\Delta X^i)^2 \ra}.
\ee

Using Schwarz inequality, we find
\be
(\d X^i)^2 (\d X^j)^2 \geq | \la \Delta X^i  \Delta X^j \ra |^2 .
\ee
Thus
\be
\d X^i \d X^j  \geq 2\pi l_s^2 \la \; |\cF^{ij}| \; \ra,   
\ee
and it generally reduces to something of the form
\be \label{uncer}
\d X^i \d X^j \geq f(g_s) \: l_s^2,
\ee
where $f(g_s)$ is some function of $g_s$. 
We will try to determine $f$ in the weak coupling limit.
Notice that this form of uncertainty relation
follows more or less from dimensional analysis.
The point of the analysis performed above is to give
a precise definition of the quantities involved
and to show that the right hand side
of (\ref{uncer}) is really nonvanishing. 

To see how the $g_s$ dependence comes in, it is instructive 
to consider the case of a D-string.
The commutation relation for the worldsheet directions reads
\be \label{xx-flat}
 [X^0, X^1] = 2\pi i l_s^2 {\cF}, 
\ee
where 
$\cF=\cF_{01}$ and we have replaced $g_{ij}$ by the flat metric in our
approximation.
We thus need to evaluate 
\be
\frac{1}{Z}\int[D\cB] e^{-S} |\cF| .
\ee
For small $g_s$, we can use the tree level SUGRA action
where $S$ contains a piece 
\be
S= \frac{1}{g_s^2 l_s^8} \int dB*dB+ \cdots .
\ee
It is convenient to go to a gauge in which $\cF=B$ and by rescaling 
\be
B = g_s B',
\ee
then for the background $B =0$
\bea
\la \; |\cF| \; \ra 
&\simeq&\frac{1}{Z_B} \int[DB] e^{-\frac{1}{g_s^2 l_s^8} \int 
( \del B_{01})^2 } |B_{01}|  \nn\\
&=& g_s
\{\frac{1}{Z'_B} 
\int[DB']e^{-\frac{1}{l_s^8}
\int (\del B'_{01})^2} |B'_{01}| \}, \label{cF}
\eea
where
\be
Z_B=\int[DB]e^{-\frac{1}{g_s^2 l_s^8}\int(\del B_{01})^2}, \quad\quad
Z'_B=\int[DB']e^{-\frac{1}{l_s^8}\int(\del B'_{01})^2}.
\ee
The above term 
$\{\cdot\}$ in (\ref{cF}) is independent of $g_s$.
One can further scale $B'_{01}$ by $l_s^4$ so that (\ref{cF})
reads $\la\;|\cF|\;\ra \simeq g_s l_s^4 I$, where 
$I$ is a path integral with no
apparent dependence on $g_s$ or $l_s$. Because of the absolute sign in
\eq{cF}, it is easy to show that $I$ is non-vanishing and is in fact
divergent. A momentum cutoff at
$\Lambda$ has to be introduced to make sense of $I$
and one finds $I \sim \Lambda^4$ and hence 
%
\be \label{D1-0}
\la\;\cF\;\ra\simeq g_s l_s^4\Lambda^4.
\ee
The natural cutoff here is $\Lambda\sim l_s^{-1}$ because we have
ignored all stringy corrections of higher order in $\alpha'$
in the SUGRA action.

While the derivation above is not completely rigorous, we consider
the possibility of extending this result consistently to all
other branes and strings via string dualities in the rest of
this paper as a supporting evidence for (\ref{D1-0}).

In principle, there could also be other sources contributing to the 
uncertainty other than the NS-NS 2-form gauge field. Here we
considered only the massless mode of SUGRA. 
In an approximation better than \eq{xx-flat},
quantum fluctuations of the metric also contribute. 
One should also take into account string loop effects
for a generic $g_s$. These can enter in at least two ways: 
First, eqn.(\ref{xx1}) was derived from a single string
in the first order approximation;
in general higher order terms and the string loop effects
can modify the commutation
relations for the D-brane worldvolume coordinates \cite{Schom}. 
Second, more precisely one should also use the string loop corrected
SUGRA action instead of the tree level one in the above derivation. 
Including all these factors, 
we expect the uncertainty relation to take the generic form
\footnote{
Uncertainty relations for the D-brane worldvolume has also been discussed
in the context of Liouville string theory \cite{ellis}. There an
uncertainty relation of a form similar to \eq{D1-1} was found, but
with a  different dependence ($\sqrt{g_s}$ instead of $g_s$) on the
string coupling.
}
\be \label{D1-1}
\d X^0 \d X^1 \geq g_s l_s^2 +\cdots, 
\ee
up to a numerical factor which will be ignored in this paper; and 
the omitted terms are of higher order in $g_s$ and $\a'$
due to the above-mentioned higher order corrections.
There could also be a dependence on $\Psi$ in the higher order
corrections.
Obviously one can perform the same derivation for a D$p$-brane with $p>1$
and find the same uncertainty relation between any two directions
on the D-brane. 

At this point, one may ask a number of questions.
For example, is it possible to determine explicitly the higher order 
corrections in \eq{D1-1}?
How does the above generalize to the case of the other D$p$-brane?
How does the uncertainty relation look like on a D$p$-brane?
What we will do is to find new uncertainty relations by
requiring the uncertainty relations to be consistent with the known
dualities of string theory. This consistency requirement  
will be our guiding principle.

\section{Worldvolume uncertainty relations}\label{ur}

In this section, we will propose a form of the uncertainty relations
for D$p$-branes 
which is consistent with
dualities of string theory. To strengthen the starting ground
for our argument which leads to the general result,
we first consider the cases of D1 and D0-branes in more detail.

\subsection{D1-branes}

{}From sec.\ref{noncomm}, we find that the uncertainty relation for a
D-string in the small $g_s$ limit in flat spacetime takes the form 
\be \label{D1}
\d X^0 \d X^1 \geq g_s l_s^2.
\ee 
Eqn.(\ref{D1}) gives a minimal area
for the D1-brane worldsheet.

An independent support for this result can be obtained by S-duality.
Under S-duality, a D1-brane is interchanged with a fundamental string,
and the string tension interchanged with the D1-brane tension.
Thus the uncertainty relation for a fundamental string should be
\be \label{F1}
\d X^0 \d X^1 \geq l_s^2.
\ee
This is in fact what people have suggested 
based on 
properties of string scattering amplitudes; 
worldsheet conformal invariance
and other various arguments \cite{Yon1}.
This can also be heuristically argued as follows 
(first two references of \cite{Yon1}). 
According to the canonical uncertainty relation in quantum mechanics
\be \label{Et}
\d E\d T\geq 1,
\ee
(where $T$ should be identified with $X^0$,)
if $\d T$ is small, $\d E$ will be large.
Since $E\sim 1/\a'$ times string length,
it is associated with a large uncertainty
$\d X^1$ in the string length \cite{DKPS}.
(For this argument to be more rigorous, we need a virial theorem
stating that a certain portion of the energy must be attributed
to the potential energy due to string tension.)

\subsection{D0-branes} \label{D0-brane}

The usual uncertainty principle of quantum mechanics
(\ref{Et}) implies that
\be
\d T\geq\frac{1}{\d E}\geq\frac{1}{E},
\ee
where we assumed that $E\geq\d E$.
(This would be the case if $E$ is positive definite.)
In the rest frame of the D0-brane, $E=1/(g_s l_s)$
is the mass of a D0-brane, so we find
\be \label{D0}
\d T\geq g_s l_s,
\ee
where $T$ is the proper time for the D0-brane 
worldline.

We can also interpret (\ref{D0}) as follows.
If (\ref{D0}) is not satisfied, the energy of a D0-brane
can be larger than its rest mass and can thus lead to 
pair productions of D0 and anti-D0-branes.
One can then imagine a quantum path in which
the created anti-D0-brane annihilates the original D0-brane
so that the created D0-brane survives as the final D0-brane.
In such cases the proper time is ill-defined during
the process of creation and annihilation.

In \cite{DKPS} it was explained that the short distance behavior
of D-branes is described by the low energy physics of
open strings ending on the D-branes.
For D0-branes two different characteristic scales were found.
The first scale is the Planck scale for the 11 dimensional SUGRA,
which is $g_s^{1/3}l_s$ in accordance with the duality between
M theory and type \IIA theory.
Assuming $g_s<1$, the Planck scale is smaller than the string
scale $l_s$ which characterizes the uncertainty relation for
fundamental strings.
The Planck scale was found as the characteristic scale of
the $0+1$ dimensional SYM theory, which describes
the low energy theory of D0-branes.
Although the Planck scale is believed to be the minimal scale
in 11D SUGRA, it is not the scale of uncertainty relations
for D0-branes since a smaller characteristic scale $g_s l_s$
was found in \cite{DKPS}.
It is called the ``fine structure'' scale, which can be seen only after
taking into account the correction of the SYM action by the DBI action.
Note that no smaller scale was found in the analysis of \cite{DKPS}.
Although the fine structure scale was obtained as the characteristic
scale in the transverse (spatial) directions,
based on our results above we propose in this paper that
the fine structure scale also sets the minimal length on
a D0-brane 
worldline as in (\ref{D0}).

Another support for the claim that (\ref{D0}) gives
the correct uncertainty relation for D0-branes is obtained via T-duality.
If the D1-brane has the uncertainty relation $\d T\d X\geq g_s l_s^2$,
then for a D1-brane wrapped on a compactified circle with radius $R$,
\footnote{
Since we assume that the 
string couplings are
smaller than $1$, the compactification radius needs to satisfy
$l_s/g'_s>R>g_s l_s$.
}
$\d X$ can not be larger than $R$, which implies that
$\d T\geq g_s l_s^2/R$.
By T-duality this is interpreted as a dual D0-brane with the uncertainty
$\d T\geq g'_s l_s$, where $g'_s=g_s l_s/R$ is the string
coupling constant in the dual theory.
This is exactly what we claimed in (\ref{D0}).

Incidentally we note that in terms of M theory,
$g_s l_s$ is the radius of the compactification
through which the M theory is dual to type \IIA theory.
By compactifying M theory on a circle smaller than the Planck scale,
the smaller scale $g_s l_s$ is introduced into the compactified M theory.
This would not be possible if there were an uncertainty relation like
$\d X\geq l_p$ in the uncompactified M theory.
On the other hand, this is consistent with
the worldvolume uncertainty relations we find for
membranes and 5-branes as in (\ref{urm2}), (\ref{urm5}).

\subsection{D$p$-branes via T-duality} \label{Td}

In the above we have seen that T-duality can be used to derive
uncertainty relation of D0-branes from that of D1-branes.
Here we will generalize the arguments to all D$p$-branes.

For simplicity we first consider the case of a flat background.
We know that a D-string can be obtained from a D2-brane under T-duality.
By wrapping a leg of the D2-brane on the circle, one introduces on the
D2-brane worldvolume an uncertainty of the order 
\be \label{X2R}
\d X^2 \sim R, 
\ee 
since the center of the D-brane can be anywhere on the circle.
It is thus natural to guess that the uncertainty relation on
a D2-brane will involve a product of uncertainties of the form
\be \label{D2}
\d X^0 \d X^1 \d X^2 \geq g_s l_s^3,
\ee
which is the product of the uncertainty relation (\ref{D1})
for a D1-brane and (\ref{X2R}) in terms of the dual $g_s$ and $l_s$.

On the other hand, one can also repeat the derivation
of D1-brane uncertainty relation in sec.\ref{noncomm} for D2-branes.
Since $\d X^i\d X^j\geq g_s l_s^2 + \cdots $ for all $i\neq j$,
$i,j=0,1,2$,  one can derive
$\d X^0\d X^1\d X^2\geq g_s^{3/2}l_s^3 +\cdots $.
This is a weaker condition than (\ref{D2}).
At this moment we do not know how to derive (\ref{D2})
directly from open strings ending on D2-branes
as in the case of D1-branes.
We leave this interesting question for future study.

For a D$p$-brane in general, we propose that
\be \label{urdp} 
\d X^0 \d X^1 \cdots \d X^p \gs g_s l_s{}^{p+1}
\ee 
in flat spacetime.
For a D$p$-brane in curved spacetime, the natural generalization
of the uncertainty relation is just
\be \label{Dpur}
\d V_{(p+1)}\geq g_s l_s^{p+1},
\ee
where $\d V_{(p+1)}$ is the uncertainty of the D$p$-brane worldvolume,
which is the spacetime volume corresponding to $\d X^i$.

One can check that \eq{urdp} respects the T-duality of
string theory. To see this, suppose that we start with a D$p$-brane in
a string theory compactified on a circle of radius $R$ with string
coupling $g_s$. We take one of the dimensions of  the D$p$-brane, says
$X^p$, to be wrapped on the circle. Since $\d X^p \sim R$, we get for
the D$(p-1)$-brane,  
\be 
\d X^0 \d X^1 \cdots \d X^{p-1} \gs g_s' l_s{}^p, 
\ee 
where
$g_s' = g_s l_s/ R$ is the dual string coupling.  Thus \eq{urdp} is
consistent with T-duality.

It is interesting to note that \eq{urdp} can be derived from
(\ref{Et}) and
\be \label{ev} 
\d E = \cT \d V_p,
\ee
where $\cT =1/ \; g_s l_s^{p+1}$ is the tension of the D$p$-brane. 
Consider an experiment conducted on the D$p$-brane which 
is supposed to measure some point-like process. Because of the
nonvanishing spacetime uncertainty relation,
the region of the process would
appear to have a spatial volume  uncertainty of 
order $\d V_p$.  The associated uncertainty in
energy would have a typical order of \eq{ev} and would be
consistent with the standard $T-E$ type uncertainty relation (\ref{Et}). 
However, to really  derive
\eq{urdp} from \eq{Et}, one should in principle also include other
possible sources of uncertainties
(e.g. contributions from the oscillation modes)
in \eq{ev}. The fact that one may derive \eq{urdp} by simply using
\eq{ev} suggests that there might be some sort of stringy 
virial theorem.
 
\subsection{Dyonic branes via S-duality}

S-duality is expected to be an exact symmetry of type \IIB string
theory. Under an S-duality transformation, a D$p$-brane gets transformed
into an $(m,n)$-$p$-brane for $p =1,5$. 
For the purpose of explicit illustrations,
we will spell out the $(m,n)$-string case in details.
$(m,n)$-5-branes can be treated similarly.

Starting with the uncertainty relation for a D-string (\ref{D1}),
we can derive the uncertainty relation for a $(m,n)$-string
by the $SL(2,\Z)$ transformation of S-duality. It is 
\be \label{urmnstring} 
\d X^0 \d X^1 \gs l_s^2 \frac{1}{ \sqrt{(m-n \chi)^2 +\frac{n^2}{g_s^2}}},
\ee
where $\chi$ is the axion vacuum expectation value.
Notice that the right hand side is just the inverse of the 
tension of the $(m,n)$-string. 
In particular, we get (\ref{F1}) as a special case.

\subsection{ M-branes via M/\IIB duality}

Just as the D-brane is the end-point of an open string, one can also learn
about the physics of the M5-brane by considering it as the boundary of
an M2-brane (see for example \cite{CS}  
for other applications in this direction). 
It was shown in \cite{CH} that the M5-brane
worldvolume can also become noncommutative in the presence of
$\cF_{ijk}$,
which is a combination of the three-form gauge field
and a worldvolume field strength.

Applying similar considerations as in the string case above, we
propose the following \ur for the M2-brane and M5-brane, 
\bea
&\d X^0 \d X^1 \d X^2 \gs l_p^3
\quad \mbox{for M2-brane,} \label{urm2} \\
&\d X^0 \d X^1 \cdots \d X^5 \gs l_p^6
\quad \mbox{for M5-brane,} \label{urm5}
\eea
where $l_p$ is the 11-dimensional Planck length.
An uncertainty relation of the same form as
(\ref{urm2}) was proposed in \cite{liyon1}
as a result of (\ref{F1}) due to the M/\IIA duality,
but with a different interpretation, which is analogous
to the one they had for (\ref{uryoneya}).

We now show that \eq{urmnstring} is related to the membrane
uncertainty \eq{urm2} by using the
M/\IIB duality. It is known \cite{sch,aspin}
that \IIB string theory can be obtained from compactifying
M theory on a shrinking 2-torus with radii $R_1, R_{11}$. A
\IIB $(m,n)$-string is identified with a membrane wrapped over
the $(m,n)$-cycle on the torus with length
\be \label{lmn}
L_{(m,n)} = R_{11} \sqrt{(m-n \tau_1)^2 + n^2 \tau_2^2}.
\ee
Here $\t = \t_1 + i \t_2$ is the modular parameter of the torus and
it is identified \cite{sch,aspin} with the \IIB string theory
parameters as
\be
\t_1 =\chi, \quad\quad \t_2 = 1/g_s.
\ee

Now starting from (\ref{urm2}) and using
\be
\d X^2 \sim L_{(m,n)}
\ee
for the uncertainty for the membrane direction which is wrapped
on a cycle, we obtain immediately \eq{urmnstring}.
Here we used $l_s^2= l_p^3/ R_{11}$.

In fact, since our uncertainty relations can be obtained
mathematically from (\ref{Et}) and (\ref{ev}), the matching
of the brane spectra for dual theories implies that
the uncertainty relations must be consistent with all dualities.
For instance, it is automatically true that the M/\IIA duality
also relates (\ref{urm2}) and (\ref{urm5})
to (\ref{F1}) and (\ref{urdp}) for $p=2,4$.
Incidentally, the M/\IIA duality also gives rise to
the uncertainty relation
\be
\d X^0\cdots\d X^5\geq g_s^2 l_s^6
\ee
for the NS5-brane.

\section{Discussions} \label{remarks}

In this paper, we discussed the  uncertainty relations for the D-brane
worldvolume. We introduced the notion of a worldvolume
uncertainty and explained how it is defined  within the context of
string field theory. We proposed worldvolume uncertainty relations
that are consistent with the various dualities in string theory.
We have also generalized the commutation relation for the
noncommutative gauge theory to a nontrivial background of $\cF$
in the lowest order approximation.
It would be interesting to generalize this result
to the full generality of an arbitrary background. 
This could be relevant to the interesting proposal in \cite{JR}.



In the following we remark on several related subjects.

\subsection{Comments on some other uncertainty relations}

\underline{Uncertainty relation of Wigner}

In the classical study of Wigner \cite{Wigner},
the effects of quantum mechanics on the measurability of
the spatial distance was estimated to be given by
\be \label{w1}
\d D \geq \sqrt{T/M_c}.
\ee
Here $T$ is the time scale for the process involved 
and $M_c$ is the mass of the clock. 
This analysis has recently been strengthened in \cite{Sasa},
which utilizes the existence of a Schwarzschild horizon $R_s$ 
for any massive object and thus it is necessary that
\be \label{w2}
\d D \geq R_s \sim G M_c. 
\ee 
Combining with \eq{w1}, one obtains
\be \label{w3}
(\d D)^3 \geq G T.
\ee
Notice that in this analysis,
the precision of the measurement of time is not limited.
The uncertainty relation we proposed is
consistent with these results.
For example, in the brane world
scenario, when $\d T=0$, our uncertainty relation for a D$3$-brane
says that $\d D = \infty$, which is stronger than
\eq{w1} or \eq{w3}.

The argument leading to (\ref{w3}) utilizes
the most popular reason for the belief in the existence of
spacetime uncertainty relations. That is, due to quantum mechanics
a large energy is needed to probe a small length scale,
and when the energy is too large a black hole is formed,
which forbids the measurement of distances behind the horizon.
However, in our derivation of the uncertainty relations (\ref{urdp}),
we did not mention anything about event horizon at all.
It remains to be seen how the consideration of black holes
can lead to the determination of uncertainty relations in string theory,
and whether it will lead to new uncertainty relations.

\underline{Uncertainty relation of Yoneya}

Notice that ours uncertainty relations \eq{urdp} are not 
of the same type as those proposed in
\cite{liyon1,Yon,yoneya,MS}. These authors proposed uncertainty
relations that involve the transverse coordinates while ours are solely
for the brane world. 
For example, in \cite{liyon1,Yon,yoneya}, it was proposed that
\be \label{uryoneya}
\d T \d X \gs l_s^2
\ee
for a D$p$-brane.
Here $\d T$ is understood as the uncertainty in the longitudinal
directions on the brane (to be more precise,
$\d T=|\d\s|$, where $\s$ is the worldvolume coordinates in the static
gauge)
and $\d X$ represents the uncertainty in directions transverse to
the brane.
As it was pointed out in \cite{DKPS}, the short distance regime
of D-branes are probed by open strings. The exchange of a closed
string state between two D-branes, for example, can be viewed
as the creation and annihilation of a pair of open strings
(an open string loop diagram) due to
the modular invariance of string theory.
Therefore, the scattering of D-branes is limited by the uncertainty
of open strings, and (\ref{uryoneya}) is a direct result of (\ref{F1}).

Our uncertainty relation (\ref{urdp}), on the other hand,
is concerned with the uncertainty among longitudinal directions
on the brane, and it has a form different from (\ref{uryoneya}).
In particular ours have a form that
depends on the dimensionality of the brane and have additional
dependence on $g_s$.

\subsection{UV-IR relations, holography and uncertainty relations}

In \cite{yoneya,liyon1}, a generalized conformal symmetry was found for
the D$p$-brane super Yang-Mills action.
It is easy to check that the $(p+1)$-dimensional YM action
with coupling $g_{\YM}$, which is schematically
\be
S=\frac{1}{g_{\YM}^2}\int d^{p+1}\s((\del X)^2+X^4),
\ee
is invariant under the following scaling transformation
\bea
&X^a \rightarrow  \l X^a, \quad
\s_i \rightarrow \l^{-1} \s_i,  \label{gct1}\\
&g_{\YM}^2 \rightarrow \l^{3-p} g_{\YM}^2. \label{gct2}
\eea
Here $X^a (a= p+1, \cdots, 9)$ are the transverse scalars, $\s_i
(i=0,1,\cdots p)$ are the worldvolume coordinates in the static gauge.
The coupling $g_{\YM}^2$ is related to the string coupling by
\be \label{gym}
g_{\YM}^2 =g_s l_s^{p-3}.
\ee
The uncertainty relation (\ref{uryoneya}) is invariant
under this scaling \eq{gct1} together with $g_s \rightarrow \l^{3-p}
g_s$ and $l_s$ being invariant. 

Notice that \eq{gct1} is reminiscent of
the UV-IR distance relation \cite{SW,UVIR} 
in the context of AdS/CFT holography \cite{p1,p2,p3}. 
Maybe this scaling relation is a generic property for
general holographic dualities \cite{hol1,hol2}. 
It was discussed in \cite{liyon1,Min} that the
uncertainty relation \eq{uryoneya} is consistent with 
the UV-IR relation and
it was suggested that the uncertainty principle is the underlying
principle that implies the UV-IR relation, which in turn guarantees the
holographic bound \cite{hol2} to be satisfied \cite{SW}. 

Since our uncertainty relations are only involved with
the worldvolume uncertainties, the scaling (\ref{gct1})-(\ref{gct2})
does not give any nontrivial implications on our relations.
On the other hand, since our \ur implies the
existence of a minimal area, it may also be relevant
to the holographic principle and to
the verification of the holographic bound.
It would be interesting if one can see this explicitly.
Another interesting issue is that 
based on the UV-IR relation and holographic principle, it is natural
to ask what kind of spacetime property (presumably a spacetime
uncertainty relation) will be implied by the worldvolume uncertainty 
relations.  We leave these issues for future studies.

\subsection{Characteristic scale}
 
The scaling transformation (\ref{gct1})-(\ref{gct2}) can also be used
to find the characteristic scale of YM theory.
Let $\l=g_s^{\frac{1}{p-3}}$, then we scale $g_s\rightarrow 1$ and
$X\rightarrow g_s^{\frac{1}{p-3}}X$.
This means that the Higgs vacuum expectation value
$g_s^{\frac{1}{p-3}}X$ is independent of $g_s$,
and thus the YM characteristic scale is $g_s^{\frac{-1}{p-3}}l_s$.
For $p=0$ this gives the Planck scale $g_s^{1/3}l_s$.
For $p=3$ it is the string scale $l_s$.
For $p>3$ this scale is much larger than the string scale,
but for these cases YM theory is not renormalizable and
it means that we cannot trust it.

As it was mentioned in sec.\ref{D0-brane},
a characteristic scale does not have to be the minimal scale.
For the case of D0-branes, it is the characteristic scale
of the DBI action that turns out to be the minimal scale.
It is thus of interest to work out also the characteristic scale for 
the DBI theory for a D$p$-brane. 
The DBI action is
\be
S=\frac{1}{g_s l_s^{p+1}}\int d^{p+1}\s\sqrt{-\det(g+\cF)},
\ee
which is invariant under the scaling
\be
X\rightarrow \l X, \quad g_s\rightarrow \l^{p+1}g_s,
\ee
and an arbitrary scaling of $\s$.
Letting $\l=g_s^{\frac{-1}{p+1}}$, we find $g_s\rightarrow 1$
and $X\rightarrow g_s^{\frac{-1}{p+1}}X$.
This means that $g_s^{\frac{1}{p+1}}l_s$ is the characteristic
scale of the DBI action for the transverse directions.
It happens that this characteristic scale is also the one
determining the minimal volume in our uncertainty relations (\ref{Dpur}).
It is possible that this is also the minimal length scale
for the transverse directions of a D$p$-brane much like the case of
D0-branes discussed in \cite{DKPS}.

Even if both the transverse and longitudinal directions of a D0-brane
are bounded by this scale $g_s l_s$,
(\ref{uryoneya}) is still a stronger condition than just
the product of the two minimal lengths.
In sec.\ref{Td} we also mentioned that (\ref{urdp})
is a stronger condition than a product of (\ref{D1})
for each pair of longitudinal directions.
It seems that in string theory we need a complicated web of
uncertainty relations, which cannot be deduced from
a single master relation, to fully state the uncertainty property
of spacetime.

\section*{Acknowledgment}

C.S.C. thanks G. Amelino-Camelia and A. Bilal for helpful discussions.
He is also grateful to the Department of Physics and  the Center for
Theoretical Physics at the National Taiwan University for hospitality
where part of this work was carried out.
P.M.H. and Y.C.K. thank S. Das and particularly M. Li for
helpful discussions. P.M.H. thanks M. M. Sheikh-Jabbari for helpful
remarks. The work of C.S.C. is
supported by the Swiss National Science Foundation.
The work of P.M.H. and Y.C.K. is supported in part by
the National Science Council
(NSC 88-2112-M-002-042, NSC 88-2112-M-002-034)
and the Center for Theoretical Physics,
National Taiwan University, Taiwan, R.O.C.


\ed

\begin{thebibliography}{99}

\bm{douglas} M. R. Douglas, {\it Superstring Dualities, Dirichlet
	Branes and the Small Scale Structure of
    	Space}, LXIV Les Houches session on
    	`Quantum Symmetries', Aug. 1995, hep-th/9610041.


\bm{liyon1} M. Li, T. Yoneya, {\it Short-Distance Space-Time
	Structure and Black Holes in String Theory}, hep-th/9806240.

\bm{early}
	G. Venezario, Europhys. Lett. {\bf 2} (1986) 199.\\
	D. Gross and P. Mende, Nucl. Phys. {\bf B303} (1988) 407.\\
	D. Amati, M. Ciafaloni and G. Veneziano, 
	Phys. Lett. {\bf B216} (1989) 41.\\
        M. Fabbrichesi, G. Veneziano, Phys. Lett. {\bf B233} (1989) 135.\\
	K. Konishi, G. Paffuti and P. Provero, Phys. Lett. {\bf B234}
	(1990) 276.


\bm{Yon} T. Yoneya,
         {\it Schild Action and Space-Time Uncertainty Principle
         in String Theory},
         hep-th/9703078;
         {\it D-Particles, D-Instantons, and Space-Time Uncertainty
         Principle in String Theory},
         hep-th/9707002.

\bm{yoneya} A. Jevicki, T. Yoneya,
	{\it Space-Time Uncertainty Principle and Conformal Symmetry in
	D-Particle Dynamics}, Nucl. Phys. {\bf B535} (1998) 335, 
	hep-th/9805069.\\
	A. Jevicki, Y. Kazama, T. Yoneya,
	{\it Generalized Conformal Symmetry in D-Brane Matrix Models},
	Phys.Rev. {\bf D59} (1999) 066001, 
	hep-th/9810146.

\bm{bw} See for example,
	N. Arkani-Hamed, S. Dimopoulos, G. Dvali, 
	{\it Phenomenology, Astrophysics and Cosmology of Theories with
    	Sub-Millimeter Dimensions and TeV Scale Quantum Gravity}, 
	Phys. Rev. {\bf D59} (1999) 086004, hep-ph/9807344. \\
	Z. Kakushadze, S.-H. H. Tye, 
	{\it Brane World}, hep-th/9809147, 
        and references therein. 

\bm{CDS} A. Connes, M. R. Douglas, A. Schwarz, {\it Noncommutative
         Geometry and Matrix Theory: Compactification on Tori},
         J. High Energy Phys. {\bf 02} (1998) 003, hep-th/9711162.

\bm{DH} M. R. Douglas, C. Hull,
        {\it D-Branes and the Non-Commutative Torus},
        J. High Energy Phys. {\bf 2} (1998) 8, hep-th/9711165.

\bm{SJ} M. M. Sheikh-Jabbari, {\it More on Mixed Boundary Conditions
        and D-Branes Bound States}, Phys. Lett. {\bf B425} (1998) 48,
        hep-th/9712199.

\bm{CH} C. S. Chu, P. M. Ho, {\it Noncommutative Open String and
	D-brane}, hep-th/9812219.  

\bm{NCT} Y.-K. E. Cheung, M. Krogh,
         {\it Noncommutative Geometry From 0-Branes in a Background
         $B$ Field}, Nucl. Phys. {\bf B528} (1998) 185. \\
 	T. Kawano, K. Okuyama,
	 {\it Matrix Theory on Noncommutative Torus}, 
	Phys. Lett. {\bf B433} (1998) 29. \\
         P.-M. Ho,
         {\it Twisted Bundles on Quantum Torus
         and BPS States in Matrix Theory}, Phys. Lett. {\bf B434} 
	(1998) 41. \\
         See also \cite{AAS}.

\bm{AAS} F. Ardalan, H. Arfaei, M. M. Sheikh-Jabbari, {\it Mixed
         Branes and M(atrix) Theory on Noncommutative Torus},
         hep-th/9803067; {\it Noncommutative Geometry from Strings
         and Branes}, hep-th/9810072.

\bm{HV} C. Hofman, E. Verlinde,
        {\it U-Duality of Born-Infeld on the Noncommutative Two-Torus},
        hep-th/9810116.

\bm{GC} H. Garcia-Compean,
        {\it On the Deformation Quantization Description of Matrix
        Compactification}, hep-th/9804188.

\bm{ST}	See for example, M.B. Green, J.H. Schwarz, E. Witten,
	{\it Superstring Theory}, (1987) Cambridge Univ. Press. \\
	J. Polchinski, {\it String Theory}, (1998) Cambridge Univ. Press.

\bm{Schom} V. Schomerus,
           {\it D-Branes and Deformation Quantization},
           hep-th/9903205.

\bm{ellis} G. Amelino-Camelia, J. Ellis, N. E. Mavromatos,
	D. V. Nanopoulos, {\it On the Space-Time Uncertainty Relations
	of Liouville Strings and D-Branes}, 
	Mod. Phys. Lett. {\bf A12} (1997)
	2029, hep-th/9701144.


\bm{Yon1} T. Yoneya,
          p.419 in ``Wandering in the Fields'',
          eds. K. Kawarabayashi, A. Ukawa (World Scientific 1987);
          Mod. Phys. Lett. {\bf A4} (1989) 1587. \\
	M. Li, T. Yoneya, {\it D-Particle Dynamics and The
	Space-Time Uncertainty Relation}, Phys. Rev. Lett. 78 (1997) 1219,
	hep-th/9611072.


\bm{DKPS} M. R. Douglas, D. Kabat, P. Pouliot, S. H. Shenker,
        {\it D-branes and Short Distances in String Theory}, 
        Nucl. Phys. {\bf B485} (1997) 85,
        hep-th/9608024.
	

\bm{CS} C. S. Chu, E. Sezgin, {\it M-Fivebrane from the Open
  	Supermembrane}, JHEP {\bf 9712} (1997) 001, hep-th/9710223. \\
	C. S. Chu, P. S. Howe, E. Sezgin, {\it Strings and D-Branes
  	with Boundaries}, Phys. Lett. {\bf B428} (1998) 59,
  	hep-th/9801202.\\
	 C. S. Chu, P. S. Howe, E. Sezgin, P. C. West, 
	{\it Open Superbranes}, Phys. Lett. {\bf B429} (1998) 273,
	 hep-th/9803041.

\bm{sch}  J. H. Schwarz, {\it An SL(2,Z) Multiplet of Type \IIB
	Superstrings}, Phys. Lett. {\bf B360} (1995) 13; 
	Erratum-ibid. {\bf B364} (1995) 252,
	hep-th/9508143.

\bm{aspin}  P. S. Aspinwall,  {\it Some Relationships Between Dualities
	in String Theory}, Nucl. Phys. Proc. Suppl. {\bf 46} (1996) 30, 
	hep-th/ 9508154.


\bm{JR}  A. Jevicki, S. Ramgoolam,  
	{\it Non-commutative gravity from the ADS/CFT correspondence},
	 hep-th/9902059. 

\bm{Wigner} E. P. Wigner,
            {\it Relativistic Invariance and Quantum Phenomena},
            Rev. Mod. Phys. {\bf 29} (1957) 255.

\bm{Sasa} N. Sasakura,
          {\it An Uncertainty Relation of Space-Time},
          hep-th/9903146.

\bm{MS} N. E. Mavromatos, R. J. Szabo,
	{\it Spacetime Quantization from Non-abelian D-particle
	Dynamics}, gr-qc/9807070; 
	{\it Matrix D-brane Dynamics, Logarithmic Operators and 
	Quantization of Noncommutative Spacetime}, hep-th/9808124; 
        {\it D-Branes and the Non-Commutative Structure of
        Quantum Spactime}, hep-th/9811116.

\bm{SW} L. Susskind, E. Witten, {\it The Holographic Bound in Anti-de
	Sitter Space}, hep-th/9805114. 

\bm{UVIR} C. S. Chu, P. M. Ho, Y. Y. Wu,
	{\it D-Instanton in $AdS_5$ and Instanton in $SYM_4$},
	Nucl. Phys. {\bf B541} (1999) 179, hep-th/9806103. \\
	M. Bianchi, M. B. Green, S. Kovacs, G. Rossi, 
	{\it Instantons in supersymmetric Yang-Mills 
	and D-instantons in IIB
    	superstring theory}, hep-th/9807033. \\
	A. W. Peet, J. Polchinski,
	{\it UV/IR Relations in AdS Dynamics},
	hep-th/9809022. 

\bm{p1} J. Maldacena, {\it The Large $N$ Limit of Superconformal 
	Field Theories and Supergravity}, Adv. Theor. Math. Phys. 
	{\bf 2} (1998) 231, hep-th/9711200.

\bm{p2} S.S. Gubser, I.R. Klebanov, A.M. Polyakov, 
	{\it Gauge Theory Correlators from Non-critical String Theory}, 
	Phys. Lett. {\bf B428} (1998) 105,
	hep-th/9802109.

\bm{p3} E. Witten, {\it Anti de Sitter Space and Holography},
	Adv. Theor. Math. Phys. {\bf 2} (1998) 253,
	hep-th/9802150.

\bibitem{hol1} G. 't Hooft, {\it Dimensional Reduction in 
	Quantum Gravity} , gr-qc/9310006.

\bibitem{hol2} L. Susskind, {\it The World as a Hologram}, 
	J. Math. Phys. {\bf 36} (1995) 6377, hep-th/9409089.

\bm{Min} D. Minic,
         {\it On the Space-Time Uncertainty Principle and Holography},
         hep-th/9808035.

\end{thebibliography}
